\documentclass[12pt]{iopart}
\usepackage{epsfig,graphics}

\begin{document}

\title{$\phi$-meson Production in Heavy-Ion Collisions at
    RHIC}
\author{S-L Blyth$^1$ for the STAR Collaboration}

\address{$^1$ Relativistic Nuclear Collisions Group, Lawrence Berkeley
National Laboratory, 1 Cyclotron Road, MS70R319, Berkeley, CA, 94720}

\ead{slblyth@lbl.gov}

\begin{abstract}
We present the first elliptic flow, $v_{2}$, measurement for the $\phi$-meson 
in Au+Au collisions at $\sqrt{s_{NN}}=200$ GeV. At low $p_{T}$ ($<2$ GeV/$c$), 
the $v_{2}$ is consistent with mass ordering expected from hydrodynamics, while
at higher $p_{T}$ ($>4$ GeV/$c$), it follows the same trend as the $K^{0}_{S}$
$v_{2}$ and a parameterization for number of quarks = 2. The nuclear modification factor,
$R_{CP}$, has also been measured and it follows the same trend as the $K^{0}_{S}$ rather than
the $\Lambda$ which confirms the baryon/meson-type dependence of $R_{CP}$ at RHIC. 
A model based on the recombination of thermal $s$-quarks
describes the central $\phi$-meson $p_{T}$ spectrum as well as the
baryon/meson ratio of $\Omega/\phi$ up to   
$p_{T}\sim4$ GeV/$c$. 
\end{abstract}  


\section{Introduction}
The hot and dense medium created in heavy-ion collisions at the Relativistic 
Heavy Ion Collider (RHIC) is extremely short-lived which requires  
probes from the early stage of the collision
to extract information about the medium and its constituents.
The $\phi$-meson, which is the lightest bound state of hidden strangeness ($s\bar{s}$),
can be used as a clean probe from the early stage: the $\phi$ is assumed to have 
a small cross-section for interactions 
with non-strange particles~\cite{Shor} and it has a relatively long lifetime
($\sim$41 fm/$c$) which means it most likely decays 
outside the fireball, ensuring that its decay daughters do not lose 
information through rescattering with other hadrons. Previous STAR
measurements~\cite{STARphi200} have ruled out $K^{+}+K^{-}$ coalescence as a 
$\phi$-meson production mechanism and therefore the information carried by 
the $\phi$ should remain unmodified by the hadronic phase.

The elliptic flow parameter, $v_{2}$, is an observable which provides
information from the early stage of the system's evolution.
Information concerning 
the partonic collectivity of the medium can be extracted from $v_{2}$ measurements 
of the $\phi$
by comparing its $v_{2}$ values with those of particles 
comprised of the lighter $u$ and $d$ quarks and with other multistrange hadrons 
($\Xi$ and $\Omega$).  

The $\phi$-meson can also provide information on particle production mechanisms
since it is a meson, but at the same time is almost as heavy as a $\Lambda$ baryon. 
Comparing the $\phi$-meson nuclear modification factor, $R_{CP}$, with other identified
hadrons can help to differentiate between mass-type and meson/baryon-type dependencies
of particle production.
Medium effects on particle production may be studied through comparison
of the $\phi$-meson transverse momentum, $p_{T}$, distributions with
model expectations.

Analysed data are from the Au+Au RunIV 200 GeV dataset.
For the $v_{2}$ and spectral analysis $\sim$13.5 million minimum bias events
(0-80$\%$) were analysed, while $\sim$4 million central triggered
events were used to obtain $p_{T}$ spectra in (0-5$\%$) central collisions.
The $\phi$-meson was reconstructed through its decay to two charged kaons (branching ratio
= $49.1\%$) which were identified through their ionisation energy loss in the STAR 
Time Projection Chamber (TPC). To extract the raw yield of $\phi$-mesons, an 
invariant mass, $m_{inv}$ distribution was constructed using all combinations of positively and 
negatively charged kaons from each event. The combinatorial background from uncorrelated 
$K^{+}K^{-}$ pairs was estimated using event-mixing,
scaled to the same-event distribution and then subtracted. 
The resulting $m_{inv}$ distribution was fitted with a Breit-Wigner distribution (to
describe the shape of the $\phi$ mass peak) plus a straight line (for
the residual background). 

\section{Results}
\begin{figure}[!hbt]
\begin{minipage}{0.4\textwidth}
\center
\includegraphics[width=\textwidth]{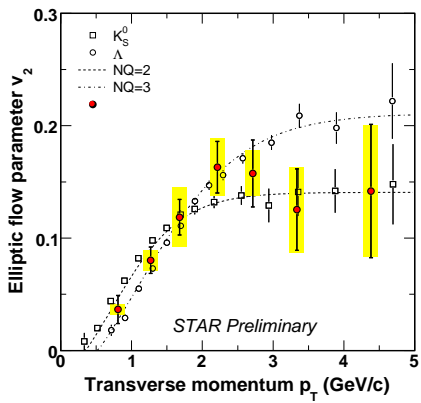}
\end{minipage}
\begin{minipage}{0.59\textwidth}
  \caption{\label{fig:v2} Elliptic flow, $v_{2}$, as a function of $p_{T}$ for the 
        $\phi$-meson (solid circles) compared to the $K^{0}_{S}$ (open squares) 
        and $\Lambda$ (open circles)~\cite{KshortLamRCP}. Statistical
        errors are represented by the error bars and the systematic errors are
        shown by the shading around the data points. The 
        dashed(dot-dashed) curve represents a parameterization for $v_{2}$ based on NQ = 2(3)
        scaling from~\cite{Xin}.}
\end{minipage}
\end{figure}

The first measurement of the $\phi$-meson elliptic flow, $v_{2}$, at
$\sqrt{s_{NN}}=200$ GeV is 
shown in Fig.~\ref{fig:v2} by the solid circles. 
The curves represent quark number (NQ) scaling $v_{2}$ 
parameterizations (dashed for NQ = 2, dot-dashed for NQ = 3)
from~\cite{Xin} which indicate the meson and baryon 
$v_{2}$ trends as a function of $p_{T}$.

The $\phi$-meson $v_{2}$ is the average of results obtained with  
two different, yet consistent, methods. The first method has been previously
established~\cite{ArtSergei} for measurements of other identified particle $v_{2}$ while the 
second method, the $v_{2}$ vs. $m_{inv}$ method, is based on~\cite{Borghini}.
The first important point to notice in Fig.~\ref{fig:v2} is that the $\phi$-meson has a 
significant $v_{2}$, which is comparable to that of other identified particles such as
the $K^{0}_{S}$ (open squares) and $\Lambda$ (open
circles)~\cite{KshortLamRCP} as well as the multistrange baryons $\Xi$
and $\Omega$~\cite{STAROmegav2} (not shown here). 
For $p_{T}<2$ GeV/$c$, the $\phi$-meson $v_{2}$ 
seems consistent with a hydrodynamic-type mass-ordering while at
intermediate $p_{T}$, ($2<p_{T}<5$ GeV/$c$), it 
is more consistent with the $v_{2}(p_{T})$ of the $K^{0}_{S}$ and the
scaling for NQ = 2, than the $\Lambda$ (NQ = 3). Therefore the $\phi$ $v_{2}$
seems to be determined by its meson properties rather than its mass. This is further evidence
of the particle species-type dependence of $v_{2}$ observed at intermediate $p_{T}$ 
at RHIC, and can be 
described in terms of particle production through quark recombination
or coalescence~\cite{Voloshin,VoloshinMolnar,FriesDuke}.

The nuclear modification factor, $R_{CP}$, can provide insight on the system created 
in central compared to peripheral collisions since it describes the yield of 
$\phi$-mesons produced in central compared to  
peripheral collisions, scaled by the number of binary collisions (calculated using
a Glauber model) and is given by
$R_{CP} = \frac{dN/dp_{T}/\langle{N_{bin}}\rangle
  |_{cent}}{dN/dp_{T}/\langle{N_{bin}}\rangle|_{periph}}$.

\begin{figure}[!hbt]
\begin{minipage}{0.39\textwidth}
\center
\includegraphics[width=\textwidth]{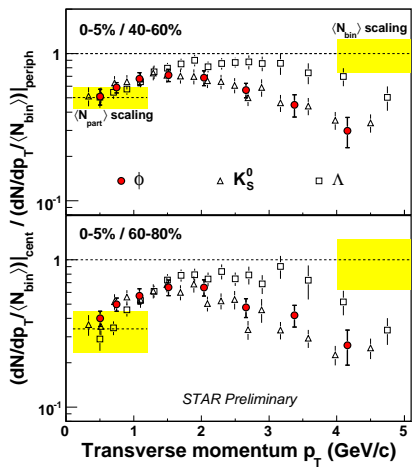}
\end{minipage}
\begin{minipage}{0.59\textwidth}
  \caption{\label{fig:Rcp} Nuclear modification factor, $R_{CP}$, for two choices of 
        peripheral bin: 40-60$\%$ (top panel) and 60-80$\%$ (bottom panel), for the 
        $\phi$-meson (solid circles). Also shown are STAR $R_{CP}$ measurements
        for $K^{0}_{S}$ (open triangles) and $\Lambda$ (open squares)
        from~\cite{KshortLamRCP}. Error bars represent statistical and systematic
        errors added in quadrature. Uncertainties due to the Glauber 
        $\langle{N_{bin}}\rangle$
        calculations are indicated by the shaded bands.}
\end{minipage}
\end{figure}  

Figure~\ref{fig:Rcp} shows $R_{CP}$ of the $\phi$-meson (filled circles) 
for two choices of peripheral
bin compared to $R_{CP}$ measurements for $K^{0}_{S}$ (open triangles) 
and $\Lambda$ (open squares)~\cite{KshortLamRCP}. 
For both choices of peripheral bin, the $\phi$ $R_{CP}(p_{T})$
is less than unity, implying that the yield in central collisions is suppressed
compared to peripheral collisions. Similarly to the trend noticed in the $v_{2}$ measurements, 
the $\phi$ $R_{CP}$ follows very closely the trend of the $K^{0}_{S}$ rather than
the $\Lambda$. This is additional confirmation of the meson/baryon-type dependence 
rather than a mass-type dependence of $R_{CP}$ at intermediate $p_{T}$ at RHIC~\cite{Kstar}. 
This grouping of mesons and baryons into separate bands in $R_{CP}(p_{T})$ has been 
predicted by recombination
and coalescence models of particle production~\cite{FriesDuke}.

\begin{figure}[!hbt]
\center
\includegraphics[scale=0.5]{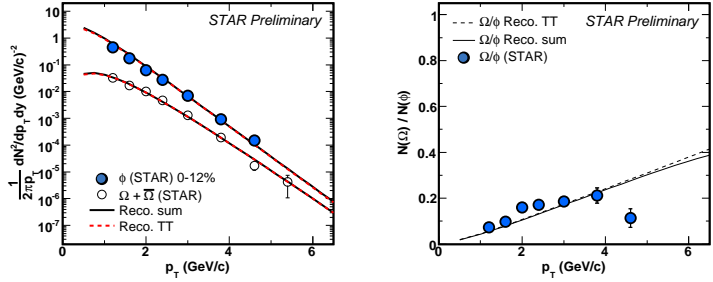}
  \caption{\label{fig:SpecRatio}Left panel shows the $\phi$ and $\Omega$ $p_{T}$ spectra 
        for the most central collisions compared to recombination model expectations 
        from~\cite{Hwa}. The sum 
        of all(thermal-thermal (TT) recombination) contributions to
        the model is shown as the solid(dashed) lines. The right panel shows the 
        $\Omega/\phi$ ratio as a function of $p_{T}$ compared to the recombination model 
        expectations. }
\end{figure}

Comparison of the $\phi$ $p_{T}$-spectra with model expectations may also provide further insight
into the constituents of the medium produced in Au+Au collisions. 
The left panel of Fig.~\ref{fig:SpecRatio} shows the central 
$\phi$-meson (solid circles) and $\Omega$ ($sss$) (open circles)
spectra compared to recombination
model expectations from~\cite{Hwa}. This model predicts that the production 
of $\phi$-mesons and $\Omega$-baryons at intermediate $p_{T}$ ($<$8 GeV/$c$) 
is predominantly due to the recombination of thermal $s$-quarks in the medium. This is
shown by the sum of all contributing recombination terms (solid lines) 
being almost identical to the term due to thermal recombination alone (dashed lines) i.e. 
the contribution to $\phi$ and $\Omega$ production from fragmentation processes 
in this model is very small. These expectations seem to match the
trend of the data.
The ratio of the $\Omega/\phi$ yields
is shown in the right panel of Fig.~\ref{fig:SpecRatio}. 
In this case, the model
predicts a monotonic increase in the $\Omega/\phi$ ratio until $p_{T}\sim8$ GeV/$c$. Although
the data deviate from the model predictions at higher $p_{T}$, their trend
can be described up to $p_{T}\sim4$ GeV/$c$.

\section{Summary and Conclusions}
In summary, the large $v_{2}$ of the $\phi$ at low $p_{T}$, despite its low
cross-section for interactions in the hadronic phase
as well as the fact that 
the $v_{2}(p_{T})$ scales with NQ = 2, are strong indications of partonic collectivity 
and deconfinement of the medium created in Au+Au collisions at RHIC. 
The good agreement of the $\phi$-meson $R_{CP}(p_{T})$ with $R_{CP}$ values of other mesons
rather than the heavier baryons, provides additional confirmation of the 
meson/baryon-type dependence of this observable at intermediate $p_{T}$.
The scaling of the $v_{2}$ and $R_{CP}$ values of the $\phi$-meson with NQ=2 is expected
from recombination and coalescence models of particle production.
In addition, the trend of the central $\Omega/\phi$ ratio can be described up to $p_{T}\sim4$ GeV/$c$
by a model based on recombination of thermal strange quarks in the medium.

\end{document}